# Theoretical investigation of the vertical dielectric screening dependence on defects for few-layered van der Waals materials†

Amit Singh,[ab] Seunghan Lee,[a] Hyeonhu Bae,[a] Jahyun Koo,[c] Li Yang[d] and Hoonkyung Lee [*a]

Cite this: RSC Adv., 2019, 9, 40309

First-principle calculations were employed to analyze the effects induced by vacancies of molybdenum (Mo) and sulfur (S) on the dielectric properties of few-layered $MoS_2$. We explored the combined effects of vacancies and dipole interactions on the dielectric properties of few-layered $MoS_2$. In the presence of dielectric screening, we investigated uniformly distributed Mo and S vacancies, and then considered the case of concentrated vacancies. Our results show that the dielectric screening remarkably depends on the distribution of vacancies owing to the polarization induced by the vacancies and on the interlayer distances. This conclusion was validated for a wide range of wide-gap semiconductors with different positions and distributions of vacancies, providing an effective and reliable method for calculating and predicting electrostatic screening of dimensionally reduced materials. We further provided a method for engineering the dielectric constant by changing the interlayer distance, tuning the number of vacancies and the distribution of vacancies in few-layered van der Waals materials for their application in nanodevices and supercapacitors.

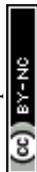

## Introduction

Dielectric screening is one of the most basic parameters of solids determining electron transfer, Coulomb interaction between electrons, charge carriers, magnetic and optical properties.[1–4] It also determines the physical, mechanical, and optical properties of van der Waals (vdW) materials, such as doping density, band gap, single photon emission, Fermi level, photoluminescence (PL), many-electron effects, and mobility carriers.[5,6] Electrostatic dielectric screening is especially important in the vertical (off-plane) direction of two-dimensional (2D) layered materials.

Exfoliation techniques are extensively used to produce various vdW 2D nanosheets,[7] including graphene,[8] molybdenum disulphide ($MoS_2$),[9] boron nitride,[10–12] metal dichalcogenides,[13] and black phosphorus (BP).[4] Due to their 2D planar structure, these materials are usually mounted on field effect transistor (FET) structures,[1–4] and they show very interesting properties, such as high performance, high carrier mobility, and high integration owing to their very small size of a few nanometers.[14] Therefore, gate switching depends on the dielectric screening of materials that induces changes in the material properties and device performance. Thus, understanding the dielectric screening in the presence of the gate field is required for predicting the optical and physical properties of these 2D materials, and for designing future nanoscale ultrathin devices.

In previous studies,[15,16] on exfoliated and CVD grown $MoS_2$, it is found that high native point defect concentration in order of $10^{13}$ cm$^{-2}$. There are many studies on $MoS_2$ related to dielectric screening,[17–19] sulphur vacancy and its interlayer interactions,[20] van der Waals interactions[21] and dynamics of defects.[22] In order to figure out the effect of vacancies, understanding of how does the gate electric field penetrates in presence of vacancies in the 2D structures and their corresponding dielectric screening in the vertical direction, is critical for designing devices and experiments in which direct measurement of the dielectric field is extremely difficult, as a result of evident quantum confinement effects and changes, that inflicts on the materials band gaps, vertical dielectric screening of vdW materials. vdW semiconductors are very sensitive with respect to the thickness of the sample; random phase approximation (RPA) and effective potential approach calculations indicate that the vertical dielectric constant varies significantly from bulk $MoS_2$ to monolayer,[23–25] and these estimations also have been widely

[a]Department of Physics, Konkuk University, Seoul 05029, Korea. E-mail: hkiee3@konkuk.ac.kr
[b]Department of Mechanical Engineering, National Chiao Tung University, Hsinchu 300, Taiwan, Republic of China
[c]Department of Condensed Matter Physics, Weizmann Institute of Science, Rehovot 7610001, Israel
[d]Department of Physics, Washington University in St. Louis, St. Louis, Missouri 63136, USA

† Electronic supplementary information (ESI) available. See DOI: 10.1039/c9ra07700f

accepted and used to explain and understand the experiments related to the capacitance measurements.[26,27]

However, recent experiments with many-layered BP, MoS$_2$, graphdiyne (GDY), and their families have shown that the dielectric constant of 2D semiconductor is not thickness-dependent,[28–30] In the weak field regime, our first-principles density functional theory (DFT) calculation based on the linear response theory suggests that the dielectric constant is thickness-independent for few-layered vdW semiconductors.[28] This result can be applied to a wide variety of vdW materials, such as SnS, GDY, MoS$_2$, and BP. Thus, in the absence of edge effects we concluded that quantum confinement does not affect the static vertical dielectric screening, it is approximately constant and independent of thickness, which implies it is the same as the bulk value. By employing the constant vertical dielectric screening and the tight binding model, the calculated band gap fluctuations of gated MoS$_2$ were found to be consistent with GW results,[28,29] which also includes many-electron effects. Therefore, based on the linear response theory, the DFT calculations are consistent and effective in predicting the dielectric properties of few-layered vdW semiconductors.

The static dielectric function, $\varepsilon(r,r')$ based on the linear response theory is given by $\delta V(r) = \int dr' \varepsilon^{-1}(r, r') \delta V_{ext}(r')$, where $\delta V$ is the calculated change in the total potential and $\delta V_{ext}$ stands for the applied external potential. The subsequent Fourier transform of the above equation gives us $\delta V(q) = \sum \varepsilon^{-1}(q, q') \delta V_{ext}(q')$. As $q' \to 0$, the expression can be reduced to $\delta V(q) = \varepsilon^{-1}(q,0) \delta V_{ext}(0)$. For bulk $q \to 0$, which resembles the dimensional average of $\delta V(r)$ over a microscopic region. The microscopic interpretation of the dielectric constant is given by the ratio of the average change of the external potential to the total potential, and it equals $\varepsilon^{-1}(0,0)$. Therefore, the dielectric constant is given by the ratio of the external to internal electric field between the layers, which gives

$$\varepsilon_r = \frac{\vec{E}_{ext}}{\langle \vec{E}_{in} \rangle} \quad (1)$$

Here, $\vec{E}_{in}$ and $\vec{E}_{ext}$ denote the internal and external electric fields, and $\langle \vec{E}_{in} \rangle$ is the mean slope calculated using a smoothed derivative of the potential between the bilayers.

The dielectric constant is a measure of the materials response to an applied electric field; especial interest is the degree of charge separation. The response to an applied electric field is a function of the electric field, state variables, frequency, temperature, pressure, mechanical stress, polarization, and surface charge density.[31] The frequency and temperature dependence of the materials dielectric properties contain useful information about transmission phenomena, defect behaviour, and structural changes. The dielectric loss and the dielectric constant values decrease as the frequency increases. This is a normal dielectric behaviour, and the basic mechanism of polarization which is similar to the conduction process. This response can be attributed to the effect of the average carrier hopping on the charge distribution of the material defects.[32] At low frequencies, the charge on the defects can be redistributed quickly, such that the defects closer to the negative side of the applied field become positively charged and vice versa. This results in the screening of the field and an overall reduction in the electric field intensity. Since the overall capacitance is inversely proportional to the field intensity, as the frequency decreases, the field intensity for a given voltage decreases, increasing the capacitance.[33] The phenomenon of atomic polarizability[34] of materials has been thoroughly studied and well established in terms of the Clausius–Mossotti relation,

$$\frac{\varepsilon_r - 1}{\varepsilon_r + 2} = \sum_i \frac{N_i \alpha_i}{3\varepsilon_0} \quad (2)$$

where $\varepsilon_r$ is the dielectric constant, $\varepsilon_0$ is the free space permittivity, $N$ is the number density of the molecules per cubic meter, and $\alpha$ is the atomic polarizability in C m$^2$ V$^{-1}$ unit.

In this work, assuming the weak field regime and using the linear response theory, we explicitly compared the external and screened electric fields, and demonstrate that screening caused by layer stacking; thus, it also includes the defect effects on the bilayer, such as Mo and S vacancy. We only considered the point defects due to their low formation energies and very likely to occur in the materials. There are many different stacking kinds of stacking in MoS$_2$ and stacking configuration are highly sensitive towards electronic properties of materials. The characteristic of vdW interlayer distance and stacking mainly influence its relative energy with respect to Γ and valence band splitting at K, as well as, the value of the optical excitations and the electron–hole binding energy.[35] The key to engineering a fundamental improved PL emission and band structure is stacking orientation in multilayer structure.[36] We choose the AA-stacking due to its most stable structure.[35,37,38] To calculate the vertical dielectric screening for MoS$_2$, we performed first-principles DFT calculations on bilayer MoS$_2$ with Mo and S vacancies.

## Computational details

All the calculations were performed using the first-principles method based on the DFT.[39] We developed and implemented our calculation based on the projector-augmented-wave method,[40] which is implemented in the Vienna ab initio Simulation Package (VASP) for realizing the Kohn–Sham eigenvalues, electrostatic potential, and wavefunctions. We employed the generalized gradient approximation (GGA) approach in the Perdew–Burke–Ernzerhof (PBE) scheme, with included vdW interactions (PBE-D2),[41,42] to describe the interlayer stacking energy (i.e., the vdW interaction). Using Grimme's approach the vdW interaction was empirically included. From our previous study,[28] the calculated vertical dielectric constant for many-layered vdW semiconductors obtained using the PBE-D2 approach agreed well with the experimental values. The Monkhorst–Pack scheme was used for the first Brillouin zone integration,[43] and the kinetic energy cutoff was chosen to be 500 eV. The 12 × 12 × 1 k-point sampling was conducted for the 2 × 2-unit cell in the first Brillouin zone to ensure that convergence is accurate to 0.001 eV per atom.

Structural optimization was performed until the Hellmann–Feynman force acting on each atom was less than 0.001 eV Å$^{-1}$.

Therefore, the calculated screening effects included the electron and ion contributions and neglected spurious interactions by the Coulomb interaction between the slabs; the vacuum gap in the vertical direction was at least 25 Å. Our calculations of vertical dielectric screening use periodic boundary conditions,[44,45] and the in-plane dimensions of $MoS_2$ were infinite. Therefore, our calculations did not account for the edge effects of $MoS_2$ on the vertical (static) dielectric screening. The interlayer (Mo–Mo) distance between AA-stacking in our calculations was 6.16 Å; this is consistent with some recent findings.[46]

## Result and discussion

We studied AA-stacked bilayer $MoS_2$, which comprises different point defects, namely the Mo and S vacancies ($V_{Mo}$ and $V_S$). We considered bilayer $MoS_2$ as displayed in Fig. 1, and illustrated the defective environment as the notation; $(D,L,N)$, where $D$ denotes the species of the defect (Mo or S), $L$ indicates the layer in which vacancies are present, and $N$ is the number of defects in the specified layer. For example, (Mo,1,4) corresponds to the case when there are four Mo vacancies in layer numbered 1 (upper layer), as shown in Fig. 1a. On the other hand, Fig. 1c shows the situation when there are eight S vacancies on the outer and inner surface of layer 1. If both layers have vacancies, it was expressed as ($D_1$–$D_2$, $L_1$–$L_2$, $N_1$–$N_2$), like (Mo–Mo,1–2,4–4) and (S–S,1–2,4–4) as depicted in Fig. 1b and d. In the previous studies,[47] it was found that up to 34% of the S can be sputtered away, S vacancies can be widely tuned with low-energy helium plasma treatment.

Vacancy concentration was defined by the ratio of the number of vacancies over the total number of atoms. It is reported that 2.56% to 8.10% of Mo vacancy[48] and 2.56% to 11.11% of S vacancy[49] can be formed. By our notation, (Mo–Mo,1–2,1–1) correspond to 5.26% of $V_{Mo}$ concentration, and (S,1,1) is equivalent to 2.56% of $V_S$ concentration.

Only bilayer $MoS_2$ systems consisting randomly distributed point defects were considered. We note that (Mo,1,2) and (Mo,2,2) are indistinguishable due to the mirror symmetry. First-principle calculations for the intrinsic bilayer $MoS_2$ was carried out firstly, and then the number of defects and distributions were varied to determine the electric properties.

To investigate the dielectric constant of the few-layered $MoS_2$, the potential differences were obtained by DFT calculation with and without the gated field, $\Delta V(z) = V_{gated}(z) - V_{ungated}(z)$. Then, the dielectric constant was given by eqn (1), based on the linear response theory. Therefore, as shown in Fig. 2b, we selected the long-wave limit ($q \to 0$), which allowed to average the potential change inside the slab.

In Fig. 2b, for bilayer $MoS_2$ with (S,1,1) vacancy, the calculated dielectric constant is 7.16. This value agrees within the range of the previously reported experimental value, 4.9–7.3.[50–52] Dielectric constants in various concentrations and distributions were also evaluated to determine the dependence of dielectric screening on defects; the results are illustrated in Fig. 2c and d. Apparently, dielectric screening strongly depends on the location and number of vacancies in the weak gate field regime. Thus, in the following, we only considered off-plane screening in the weak field regime for 2D materials. To determine whether the exotic features of dielectric screening depend on the location and number of vacancies, we calculated the dielectric constant of $MoS_2$ in the presence of S vacancies, for different configurations.

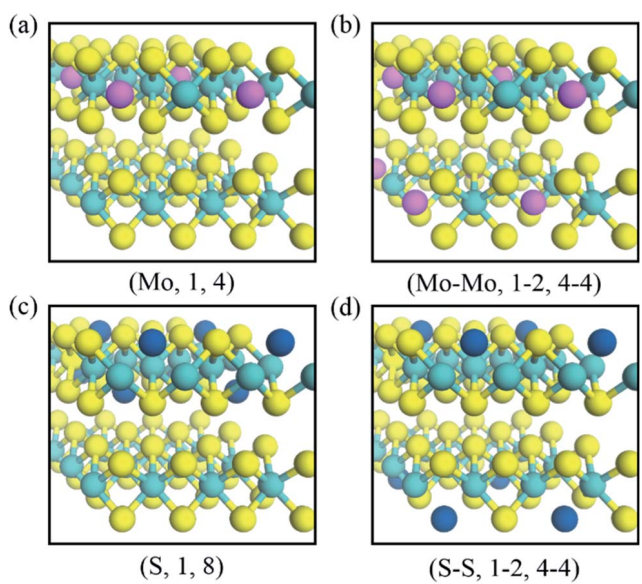

Fig. 1 Schematic of Mo and S vacancies in bilayer $MoS_2$ defined by (vacancy, layer, number of vacancies) (a) Mo vacancies in the upper layer; (Mo,1,4), (b) Mo vacancies in two layers; (Mo–Mo,1–2,4–4), (c) S vacancies in the upper layer; (S,1,8) and (d) S vacancies in two layers; (S–S,1–2,4–4).

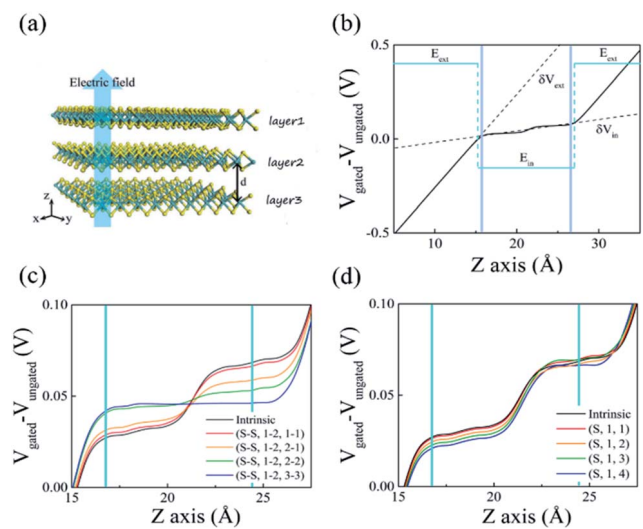

Fig. 2 (a) Atomic structure of few-layered $MoS_2$ under the gated electric field, (b) the change of the potential (solid line) in vertical direction through bilayer $MoS_2$ with S vacancy (S,1,1), under 0.5 V nm$^{-1}$ of gated field. Vertical cyan line indicated the position of $MoS_2$ layers. Vacancies are distributed: (c) when two layers contains vacancies, and (d) when only one layer contains vacancies.

Here we refer to uniformly distributed vacancy and non-uniformly distributed vacancy as the vacancies lying in one layer and both layers, respectively. Our results show that the dielectric constant increases in the case of uniformly distributed vacancies, as shown in Fig. 2c, whereas for non-uniformly distributed S vacancies, the dielectric constant decreases as presented in Fig. 2d. In Table 1, we summarized the dielectric constants of $MoS_2$ containing various S vacancies using mean slopes. The table implies that delocalized vacancies increase dielectric screening whereas non-uniformly distributed defects decrease. We further studied the atomic charge distribution under the electric field, and then employed VASP to calculate the charge density difference between the scenarios without the gate and with the gate electric field, to obtain the charge redistribution (Fig. 3). Fig. 3a and b show the change in the charge density without and with the gate electric field, for non-uniformly distributed S vacancies.

Fig. 3c and d show the change in the charge density for uniformly distributed S vacancies on bilayer $MoS_2$. As shown in Fig. 3a and c, positive charge accumulates above while negative charge accumulates below the layer, for each layer. The lower and upper layers have the highest charge density (with different polarities). Negative and positive charge accumulation is symmetrical; as we can see clearly from Fig. 3a, the polarization effect owing to the non-uniform distribution of vacancies is negligible. Therefore, the induced electric field is approximately given by $\vec{E}_{ind} = -\sigma_{ind}/\varepsilon_0 \hat{z}$, where $\hat{z}$ denotes the unit vector along the $z$-axis, and $\sigma_{ind}$ is the induced surface charge density. The net internal electric field inside is given by $\vec{E}_{in} = \vec{E}_{ext} + \vec{E}_{ind}$ ($\vec{E}_{ext} = \sigma/\varepsilon_0 \hat{z}$ and $\vec{E}_{in} = \sigma/\varepsilon_0\varepsilon_r \hat{z}$), leading to $\varepsilon_r = \sigma/(\sigma - \sigma_{ind})$. Based on the previous work, we conclude that the dielectric constant is independent of the applied electric field.[28]

In the case of non-uniformly distributed vacancies, the induced surface charge density ($\sigma_{ind}$) decreases (Fig. 3b). Based on the above discussion, we conclude that the dielectric constant decreases in the case of non-uniformly distributed S vacancies. For distributed S vacancies, the effect of polarization is significant. In this case, the dielectric constant is well-defined by the Clausius–Mossotti relation (eqn (2)), which in turn shows that the dielectric constant increases with increasing polarizability. Fig. 3c shows that the atomic polarizability $\alpha$ increases as the number of S vacancies in the uniformly distributed case. In the delocalized vacancy case, the net internal electric field is given by $\vec{E}_{in} = \vec{E}_{ext} + \vec{E}_{ind} - \vec{E}_{pol}$, where $\vec{E}_{pol}$ is an approximate field induced by the polarization. Therefore, the dielectric constant in the uniformly distributed vacancies case is obtained by

$$\varepsilon_r = \frac{\vec{E}_{ext}}{\langle \vec{E}_{ext} + \vec{E}_{ind} - \vec{E}_{pol} \rangle} \quad (3)$$

where $\vec{E}_{in}$ and $\vec{E}_{ext}$ denote the internal electric field including the polarization effect, and the external electric field, respectively. The dielectric constant is related to the electronic polarization of the material.

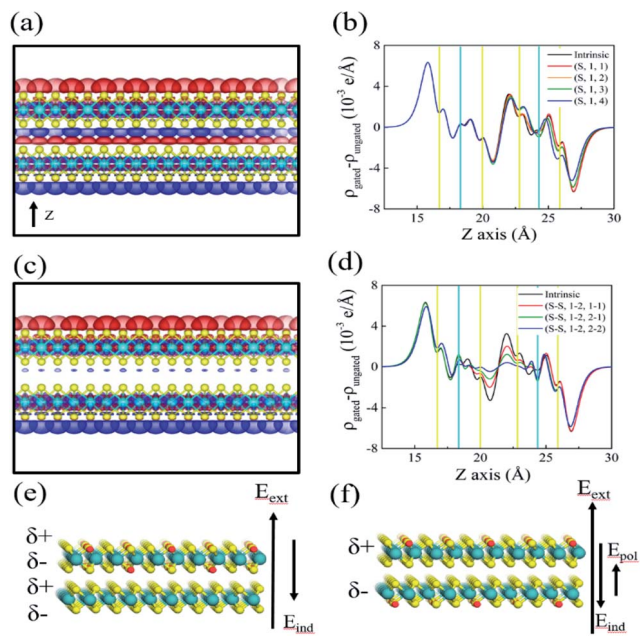

Fig. 3 Charge density difference without and with the gated electric field, when the S vacancy is (a) non-distributed and (c) uniformly distributed, respectively. The differences in charge density projected along the $z$-axis in the presence of (b) the non-distributed and (d) distributed S vacancies. The yellow and cyan vertical lines indicate the positions of S and Mo atoms, respectively. Schematic of charge accumulation and electric field in the presence of (e) the non-distributed and (f) distributed S vacancies, where red dots indicate.

Material polarization is defined as the sum of the dipole moments per unit volume, given by $P = N\alpha E$, and $\alpha$ is related to the dielectric constant by eqn (2). From the above model, $\vec{E}_{ind}$ and $\vec{E}_{pol}$ are important for determining the dielectric constants of vdW materials.

Uniform and non-uniform distributions of vacancies were considered, and it is believed that their dependence on off-plane dielectric screening can be explained by the polarizability dependence of vdW materials. In the case of non-uniformly distributed vacancies (Fig. 3a and b), the dielectric constant decreases because $\sigma_{ind}$ decreases and also polarization is negligible. On the other hand, in the case of uniformly distributed vacancies (Fig. 3c and d), the polarization effect becomes dominant and increases the overall dielectric constant of the material. Therefore, the vertical, off-plane dielectric shield is well-explained by this model, and it can be used to predict the dielectric constant and the properties of few-layered vdW materials. Fig. 3e and f schematically show the changes in the electric field induced, respectively, by non-uniform and uniform distributions of S vacancies. In the non-uniform distribution case, the charge distribution is symmetric around the centre of the bilayer which induces the polarization between the layers, as shown in Fig. 3e and f. We can see that the significant effect of polarization owing to the accumulation of charges on the upper and lower layers of bilayer $MoS_2$. Fig. 3e shows the schematic of $\vec{E}_{pol}$ effects on the internal electric field.

Table 1 The calculated dielectric constant of $MoS_2$ with various S vacancies based on DFT calculation. We obtained the mean slopes and calculated dielectric constant using eqn (1). Yes or no for polarization dominance indicate dominant and non-dominant contributions of the field by the polarization, respectively

| Sulphur vacancy configuration | Dielectric constant | Polarization dominance |
|---|---|---|
| Intrinsic | 7.00 | No |
| (S,1,1) | 7.16 | No |
| (S–S,1–2,1–1) | 8.11 | Yes |
| (S,1,2) | 7.12 | No |
| (S–S,1–2,1–2) | 10.5 | Yes |
| (S,1,3) | 6.85 | No |
| (S–S,1–2,2–2) | 19.6 | Yes |
| (S,1,4) | 6.70 | No |

Next, the effect of Mo vacancies on the vertical dielectric screening of bilayer $MoS_2$ was considered. We obtained the charge densities without and with the gated electric field (Fig. 4b) and plotted the redistribution of charge (Fig. 4c and d). As shown in Fig. 4a, $\vec{E}_{in}$ decreases in both case with increasing the number of vacancies, which is totally different from the case of S vacancies. In the case of Mo vacancies, the dielectric constant increases for both uniformly and non-uniformly distributed vacancies. The $MoS_2$ layer consists of Mo that is stacked by two layers of S, which in turn creates more polarization even in the case of non-uniform distribution (Fig. 4c); therefore, charge accumulates at the lower and upper layers. In the case of uniformly distributed vacancies, the effect of polarization is higher (Fig. 4d) than in the non-uniform distribution case. The dielectric constants for Mo vacancies were calculated using eqn (1) and concluded that in the Mo vacancies case the dielectric constant always increases. Therefore, for the both Mo vacancy scenarios, an approximate dielectric constant is given by eqn (3). Careful investigation of potential profiles in Fig. 2b–d establishes that the potential is much less screened in the interlayer and is robustly screened in the layers. Therefore, we can tune the dielectric constant by tuning the interlayer distance; consequently, vertical dielectric screening can be effectively designed and controlled in vdW materials.

As shown in Fig. 5, some cases of uniformly distributed and non-uniformly distributed vacancies, namely (Mo–Mo,1–2,1–1) (Fig. 5b), (S–S,1–2,1–1) (Fig. 5c) and (S,1,2) (Fig. 5d) were selected and interlayer distances (*i.e.*, the perpendicular distance between Mo–Mo atoms in adjacent layers) were varied manually. It was found that off-plane dielectric screening strongly depends on the interlayer distance. We further investigated the dielectric constant for all the cases, including S and Mo vacancies for both the uniform distribution and non-uniform distribution cases, and obtained the potential difference between the scenarios with and without the gated field, to determine the dependence of the dielectric constant on the interlayer distance. We employed eqn (1) to calculate the dielectric constant. As shown in Fig. 5a–d, the internal electric field decreases as the interlayer distance decreases, which increases the dielectric constant. These results are in a good agreement with our previously reported values.[28]

To investigate the change in the dielectric screening of $MoS_2$ in the presence of heterogeneous vacancies, we employed Mo and S vacancies and used two different types of models. First, we used a (S–Mo,1–2,1–1) distributed vacancy; next, we used a (S–Mo,1,1–1) vacancies on vertical dielectric screening (Fig. 6a). We concluded that in the hetero-vacancy case, off-plane dielectric screening can be explained by the effect of polarizability given by eqn (2), and the approximate dielectric constant varies according to eqn (3). To demonstrate the

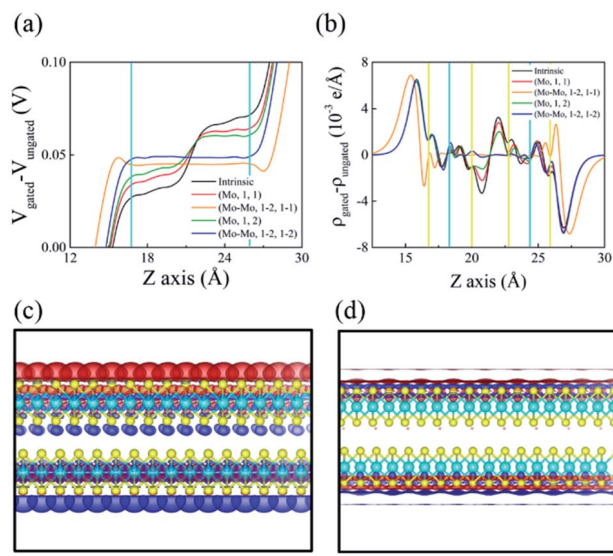

Fig. 4 (a) Potential difference according to the position and location of the Mo vacancy for bilayer $MoS_2$. (b) The difference in the induced charge density projected along the *z*-axis. (c) Charge densities, for the non-distributed Mo vacancy case, (Mo–Mo,1–2,1–1) and (d) distributed Mo vacancy case, (Mo,1,2).

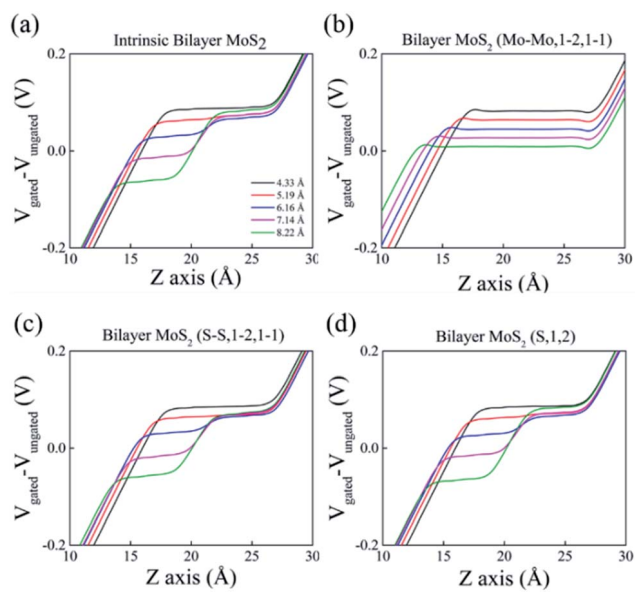

Fig. 5 Potential change $\Delta V(z) = V_{gated}(z) - V_{ungated}(z)$, calculated along the *z*-axis to the calculated change in the dielectric constant of bilayer $MoS_2$ for (a) the intrinsic, (b) Mo vacancy (Mo–Mo,1–2,1–1), (c) S vacancy (S–S,1–2,1–1), and (d) (S,1,2) cases, for various interlayer distances.

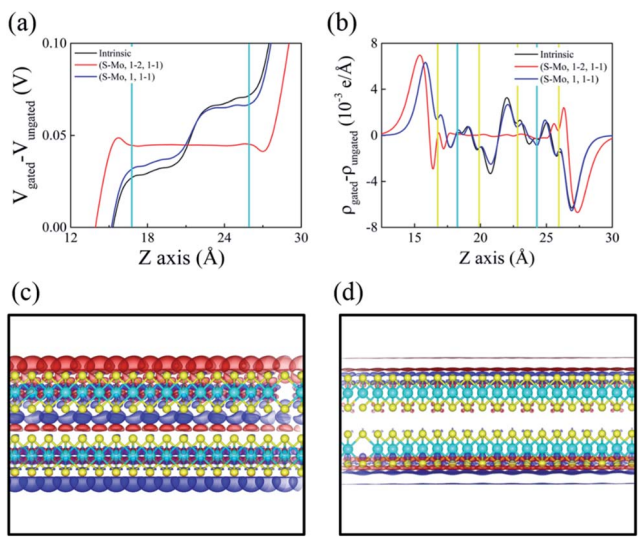

Fig. 6 (a) Potential difference, computed in the hetero-vacancy case, with and without the gated field projected along the z-axis. (b) Induced charge density projected along the z-axis. (c) Representation of charge density for non-uniformly distributed vacancy case (S–Mo,1,1–1). (d) Representation of charge density for the uniformly distributed vacancy case (S–Mo,1–2,1–1).

dependence, we calculated the potential difference without and with the gated field, as shown in Fig. 6a, which shows that the dielectric constant increases in both the uniformly distributed and non-uniformly distributed cases, where we used eqn (1) to compute the dielectric constant. To explain the dependence, we further investigated the microscopic charge distribution and the difference between the charge densities without and with the gated field, to obtain the charge redistribution, as shown in Fig. 6b. The charge accumulated at the upper and lower layers, as illustrated in Fig. 6c and d, which induced polarization and increased the dielectric constant. Fig. 6c shows the charge redistribution for the non-uniformly distributed vacancy case (S–Mo,1,1–1), in which we can see that the polarization effect is small, compared with that for the uniformly distributed case (S–Mo,1–2,1–1), shown in Fig. 6d, which is consistent with the results for the uniformly distributed S vacancies. Therefore, the dielectric constant of bilayer $MoS_2$ in the hetero-vacancy case can be calculated by considering the effect of the materials polarizability and using eqn (3), which is consistent with the linear response theory.

## Conclusions

In conclusion, based on the linear response theory and DFT calculations, we investigated the dependence of vertical dielectric screening of $MoS_2$ on point defects, such as Mo and S vacancies. By exploring the uniformly and non-uniformly distributed vacancy cases of S and Mo for vdW bilayer $MoS_2$, we concluded the following: (1) the dielectric constant of $MoS_2$ with point defects remarkably depends on the distribution of vacancies, because of the polarization effect. In the case of non-uniformly distributed S vacancies, the dielectric constant decreases. On the other hand, for uniformly distributed S vacancies, the dielectric constant increases due to the interlayer polarization. For Mo vacancies, the dielectric constant increases both for uniformly and non-uniformly distributed cases, because Mo linkage with neighbouring atoms induces comparable polarization. (2) The dielectric constant strongly depends on the vacancy concentration, even in the small concentration we can see the significant changes on the dielectric properties. (3) The dielectric constant for hetero-vacancies exhibits the same behaviour as in the Mo vacancies case, owing to the dominant effect of Mo vacancies, which in turn increases the overall effect of polarization; therefore, the dielectric constant increases in both cases. Out model can be widely used for understanding how vacancies affect the electronic properties of vdW materials and the off-plane (vertical) dielectric screening of 2D materials.

From the application point of view, our results can be used in nanocapacitors engineering for energy storage, and dielectric constants of supercapacitors. The dielectric constant of 2D materials can be widely adjusted by controlling the type of vacancy, interlayer distance, and distribution of vacancies. On the other hand, we can also investigate the location of vacancies (defects) along the z-axis, by comparing intrinsic behaviours. Using this, the distribution of defects can be deduced from observed off-plane dielectric constant of vdW materials. We proposed a method for developing vdW materials whose dielectric constants can be widely tuned and engineered, which is advantageous for future applications in the fields of nanoscience and nanotechnology.

## Conflicts of interest

The authors declare no competing financial interest.

## Acknowledgements

This work was supported by the Basic Science Research Program (Grant No. KRF-2018R1D1A1B07046751 and NRF-2017M3C1B6070572) through the National Research Foundation (NRF) of Korea, funded by the Ministry of Education, Science and Technology. This paper was written as part of Konkuk University's research support program for its faculty on sabbatical leave in 2018.